\documentclass[showpacs,floatfix,superscriptaddres, twocolumn]{revtex4-1}
\usepackage{graphicx,psfrag,amsmath,amssymb,amsfonts,latexsym,color,epsf,graphpap,dcolumn}
\usepackage{graphicx}
\usepackage{caption}
\usepackage{subcaption}
\usepackage{amssymb}
\usepackage{epstopdf}
\usepackage{color, amsmath, bm}
\usepackage[dvipsnames]{xcolor}
\usepackage[hidelinks]{hyperref}
\usepackage{natbib}
\RequirePackage{amsmath}
\usepackage{amsfonts}   
\usepackage{physics} 
\usepackage{comment}
\usepackage{dirtytalk}

\date{today}
\usepackage[latin1]{inputenc}
\usepackage[T1]{fontenc} 
\usepackage[english]{babel}
\usepackage{bm}       
\usepackage{hyperref}
\pdfoutput=1

\date{today}

\usepackage[latin1]{inputenc}
\usepackage[T1]{fontenc} 
\usepackage[english]{babel}
\usepackage{bm}       
\usepackage{hyperref}
\pdfoutput=1
\begin{document}
\title{Motion induced excitation and electromagnetic radiation from an atom
facing a thin mirror}
\author{C\'esar D. Fosco $^1$ \footnote{fosco@cab.cnea.gov.ar}}
\author{Fernando C. Lombardo$^2$ \footnote{lombardo@df.uba.ar}}
\author{Francisco D. Mazzitelli$^1$ \footnote{fdmazzi@cab.cnea.gov.ar}}

\affiliation{$^1$ Centro At\'omico Bariloche and Instituto Balseiro,
Comisi\'on Nacional de Energ\'\i a At\'omica, 
R8402AGP Bariloche, Argentina}
\affiliation{$^2$ Departamento de F\'\i sica {\it Juan Jos\'e
 Giambiagi}, FCEyN UBA and IFIBA CONICET-UBA, Facultad de Ciencias Exactas y Naturales,
 Ciudad Universitaria, Pabell\' on I, 1428 Buenos Aires, Argentina}
\date{\today}
\begin{abstract}
We evaluate the probability of (de-)excitation and photon emission from a
neutral, moving, non-relativistic atom, coupled to the quantum electromagnetic
field and in the presence of a thin, perfectly conducting plane
(``mirror'').  These results extend, to a more realistic model, the ones
we had presented for a scalar model, where the would-be
electron was described by a scalar variable, coupled to an (also scalar)
vacuum field. The latter was subjected to either Dirichlet or Neumann
conditions on a plane. 
\noindent In our evaluation of the spontaneous emission rate produced when
the accelerated atom is initially in an excited state, we pay attention to
its comparison with the somewhat opposite situation, namely, an atom at
rest facing a moving mirror.	
\end{abstract} 
\maketitle
\section{Introduction}\label{sec:intro}
Several important effects are associated with the quantum vacuum
fluctuations of the electromagnetic (EM) field, ranging from the
microscopic realm (spontaneous emission by excited atoms, the Lamb shift,
anomalous magnetic moments of elementary particles, Van der Waals
interactions) to the Casimir interaction between neutral macroscopic
bodies~\cite{booksCasimir}.
The Casimir-Polder force correspond to a hybrid situation, since it
involves an atom and a macroscopic medium. Different manifestations of the
fluctuations of the EM field in analogous situations correspond to the 
effect of a change in boundary conditions on the probability of spontaneous
emission from an atom. This has been studied for an atom in the presence of a
perfectly conducting plane, or inside a cavity~\cite{cavityqed}. 

New effects arise when one introduces time dependence;  for instance, when
the atom or the macroscopic media are in motion: photon creation by
accelerated mirrors (Dynamical Casimir effect) and quantum friction for an
atom and a surface (or between two surfaces) in relative motion at constant
velocity~\cite{reviewsdce}.  
In the present work we are concerned with a dynamical situation,
focussing on the changes in the decay probability (of an initially  excited
atom) as well as on the possibility of excitation of an atom which is
initially in its ground state. Note that, for this kind of system, one can
even have the production of photon pairs without any change in the atomic
state, what is the microscopic analog of the dynamical Casimir
effect~\cite{NetoMic}.

In Ref. \cite{Fulling1}, authors considered an atom at rest near an
accelerating mirror shown that uniformly accelerated motion of the  
mirror yields excitation of a static two-level atom with simultaneous emission of a real photon. 
They also found that the excitation probability oscillates as a function of the atomic position because of interference between contributions from the waves incident on and reflected from the mirror.
In \cite{Fulling2} an atom
accelerating near a mirror is considered and a radiative effect is reported. From an inertial point of view, the process arises from 
a collision of the negative vacuum energy of Rindler space with the mirror. There is a qualitative symmetry under interchange of 
accelerated and inertial subsystems, but it hinges on the accelerated detector's being initially in its own Rindler vacuum.

In a previous work~\cite{PaperI},  we presented a study on the excitation
and decay probabilities for a moving atom in front of a planar mirror, in a
simplified model: the ``atom'' was endowed with a scalar variable describing
the electron, and it was coupled to a real quantum scalar field. Perfect
conductor boundary conditions were replaced with Neumann and Dirichlet
boundary conditions on the mirror's plane.

We paid particular attention to two different processes, both taking place
to the lowest order in the coupling constant:  transition of the atom
from the ground state to the first excited state, with simultaneous
emission of a photon, and spontaneous emission of an initially excited
atom. We considered a small-amplitude motion of the atom,  and analysed the
spectral and directional dependence of the radiation on the motion. 

In this paper we generalise those results to the more realistic  case of an
atom coupled to the quantum electromagnetic field, taking into account the 
$\mathbf v \times\mathbf B$ interaction between the moving atom and the
magnetic field, the so called R\"ontgen term, which is a consequence of the
vector character of the electromagnetic field and is crucial to maintain
Lorentz covariance~\cite{Wilkens} to the relevant order in the velocity.

This paper is organized as follows. In Sect.~\ref{sec:thesystem} we
describe our model in terms of its classical action. In
Sect.~\ref{sec:effective_action} we compute the vacuum persistence
amplitude from the imaginary part of the effective action. After recovering
known results for an atom oscillating in free space, we also present the
calculation for the case of the atom oscillating in front of a planar
perfect mirror.  In Sect.~\ref{sec:amplitudes}, we compute the transition
probabilities for two processes: decay of an excited atom and excitation of
an atom initially in its ground state. We compare the results for a moving
atom in front of a static mirror with those for  an oscillating mirror and
a static atom~\cite{Zoller, Passante}, tracing the differences in terms of
the R\"ontgen current.  Section~\ref{sec:conc} contains the conclusions of
our work.  
\section{The model and its classical action}\label{sec:thesystem}
Our starting point to define the model, shall be the action ${\mathcal
S}_a$, for an atom coupled to the EM field, in the electric dipole
approximation.  This approximation, to be unambiguous, must be formulated
on a comoving system.  Indeed, let us assume that in the system where the
atom is at rest there is an electric dipole moment ${\mathbf d}_0$, and a
vanishing magnetic dipole: ${\mathbf m}_0 = 0$. Then, in the Lab system,
and to the first order in the velocity of the atom (in our conventions, the
speed of light $c\equiv 1$) we shall have: 
\begin{align} 
{\mathbf d}(t) & = {\mathbf d}_0(t)+ {\mathbf v}(t) \times 
 {\mathbf m}_0(t) = {\mathbf d}_0(t) + {\mathbf v}(t) 
 \times \, 0 =  {\mathbf d}_0(t) \nonumber\\
{\mathbf m}(t) & = {\mathbf m}_0(t) \,-\, {\mathbf v}(t) \times
 {\mathbf d}_0(t) =- {\mathbf v}(t)  \times  {\mathbf d}_0(t)  \nonumber\\
 & = - {\mathbf v}(t)  \times  {\mathbf d}(t).
\end{align}
Therefore, the action ${\mathcal S}_a$ (in the Lab system) must also
include a coupling, to the magnetic field, of the motion-induced magnetic
dipole:
\begin{align}
{\mathcal S}_a & =\; \int dt \, 
	\Big[ \frac{m}{2} \dot{\mathbf x}^2(t) \,-\, V({\mathbf x}(t))
\,+\, {\mathbf d}(t) \cdot {\mathbf E}(t,{\mathbf r}(t)) \nonumber \\
&\,+\, {\mathbf m}(t) \cdot {\mathbf B}(t,{\mathbf r}(t))
\Big] \nonumber\\
& =\; \int dt \, 
	\Big\{ \frac{m}{2} \dot{\mathbf x}^2(t) \,-\, V({\mathbf x}(t))
	\nonumber \\ &\,+\, e \, {\mathbf x}(t) \cdot \big[ {\mathbf E}(t,{\mathbf r}(t)) 
\,+\, \dot{\mathbf r}(t) \times 
{\mathbf B}(t,{\mathbf r}(t)) \big] \Big\}\;, \label{2}
\end{align}
where ${\mathbf x}(t)$ denotes the position of the electron with respect to
the (center of mass of the) atom while ${\mathbf r}(t)$ does so for the
atom with respect to the origin of the Lab system.  Regarding the potential
$V$ binding the electron, for the sake of simplicity,  we shall assume here
that it has a harmonic oscillator form:
\mbox{$V = \frac{m}{2} \Omega^2 {\mathbf x}^2$}. 

On the other hand, the free EM field action ${\mathcal S}_{\rm em}(A)$ is
given by
\begin{equation}\label{eq:defsg0}
{\mathcal S}_{\rm em}(A) \;=\; \int d^4x \, \left[-\frac{1}{4}  F_{\mu\nu}
F^{\mu\nu} \,+\, {\mathcal L}_{\rm g.f.}(A) \right] \;,
\end{equation}
with $F_{\mu\nu} = \partial_\mu A_\nu - \partial_\nu A_\mu$, which includes
a gauge-fixing term ${\mathcal L}_{\rm g.f.}$.  We want
to consider the cases of an atom moving either in free space or in the
presence of a perfect mirror. The second case shall be dealt with when
integrating out the EM field fluctuations. Not unexpectedly, the outcome
will turn out to be the sum of the free space result plus a ``reflected''
contribution (in the method of images sense).

\section{Effective action and its imaginary part}\label{sec:effective_action}

As a first step in the derivation of the effective action $\Gamma[{\mathbf
r}(t)]$, which will only depend on the atom's trajectory, we  first
integrate out the electron's degrees of freedom, ${\mathbf x}(t)$, to
obtain an intermediate effective action 
${\mathcal S}_{\rm eff}(A; {\mathbf r})$.

Since we are assuming a harmonic oscillator form for $V$ in (\ref{2}), the functional integral over 
${\mathbf x}$ becomes a Gaussian. The result of such an integral is (modulo an irrelevant constant) tantamount to replacing
{\em in the action\/} the integrated variable in terms of its source, using the classical equation of motion for ${\mathbf x}$. 
The latter corresponds to a harmonic oscillator forced by a time-dependent force, which is the Lorentz force acting on ${\mathbf r}(t)$ (not on ${\mathbf x}$) .
Solving for ${\mathbf x}$ in terms of that force, and recalling that Feynman conditions are to be imposed on the time dependence of that solution, we find:

\begin{equation}
{\mathcal S}_{\rm eff}(A; {\mathbf r}) = {\mathcal S}_{\rm em}(A) + {\mathcal
S}_{\rm I}^{(a)}(A; {\mathbf r}) \;,
\end{equation}
with
\begin{align}
&	{\mathcal  S}_{\rm I}^{(a)}(A, {\mathbf r}) =  \frac{i e^2}{2 m}
	 \int_{t,t'}  \Delta_\Omega(t-t')  
	\big[ {\mathbf E}(t,{\mathbf r}(t)) 
 \\ & + \dot{\mathbf r}(t) \times 
{\mathbf B}(t,{\mathbf r}(t)) \big]  \cdot 
\big[ {\mathbf E}(t',{\mathbf r}(t')) 
+\dot{\mathbf r}(t') \times 
{\mathbf B}(t',{\mathbf r}(t')) \big]  \nonumber 
\end{align}
where we have used a shorthand notation for the integration over time, for example: 
\mbox{$\int_{t,t'} \ldots \equiv \int_{-\infty}^{+\infty} dt 
\int_{-\infty}^{+\infty} dt' \ldots$,} and: 

\begin{eqnarray}\label{eq:defdelta}
\Delta_\Omega (t-t') & =&  \int \frac{d\nu}{2\pi} 
e^{-i \nu (t-t')}  \widetilde{\Delta}_\Omega(\nu) \nonumber \\ 
\widetilde{\Delta}_\Omega(\nu) &=&
\frac{i}{\nu^2 - \Omega^2 + i \epsilon} .
\end{eqnarray}
We can produce an explicit expression for $\Delta_\Omega (t-t')$, which will turn out to be quite useful:
\begin{equation}\label{eq:deltaexp}
\Delta_\Omega (t-t') \;=\; \frac{1}{2 \Omega} \, 
\big[ \theta(t-t') \, e^{-i \Omega (t-t')} 
\,+\, \theta(t'-t) \, e^{ i \Omega (t-t')} 
\big] \;.
\end{equation}

The final form of the effective action of the system, $\Gamma[{\mathbf
r}(t)]$, is obtained  by including the  EM field fluctuations. Namely,
\begin{equation}\label{eq:gammar}
e^{i \Gamma[{\mathbf r}(t)]} \;=\;
\frac{\int {\mathcal D}A \, 
e^{ i {\mathcal S}_{\rm eff}(A; {\mathbf r})}}{\int {\mathcal D}A \, e^{ i {\mathcal S}_{\rm eff}(A; {\mathbf r}_0 )}} \;,
\end{equation}
where ${\mathbf r}_0$ is the average position of the atom, which we will assume to be time independent. 
Then, up to first order in $e^2$ we obtain

\begin{align}\label{Gammaam}
	\Gamma[{\mathbf r}(t)] &=\; \frac{i e^2}{2 m} \;
	\int_{t,t'} \; \Delta_\Omega(t-t') \,
	\Big[ \big\langle {\mathbf E}(t,{\mathbf r}(t)) \cdot 
	{\mathbf E}(t',{\mathbf r}(t'))  \rangle \nonumber\\
	& +\, 2 \, \big\langle {\mathbf E}(t,{\mathbf r}(t)) 
\cdot \dot{\mathbf r}(t') \times 
{\mathbf B}(t',{\mathbf r}(t')) \big\rangle
\nonumber \\  &+\, \, \big\langle \dot{\mathbf r}(t) \times 
{\mathbf B}(t,{\mathbf r}(t))
\cdot \dot{\mathbf r}(t') \times 
{\mathbf B}(t',{\mathbf r}(t')) \big\rangle
  \nonumber\\
  & -\, \big\langle {\mathbf E}(t,{\mathbf r}_0) \cdot 
	{\mathbf E}(t',{\mathbf r}_0)  \rangle \Big] \;, 
\end{align}
where the symbol $\big\langle \ldots \big\rangle$ denotes the functional
averaging
\begin{equation}\label{eq:eecor}
\big\langle \ldots \big\rangle \;=\; \frac{\int {\mathcal D}A \;  \ldots \; \exp \Big\{ i \big[
{\mathcal S}_{\rm em}(A) \big]\Big\}}{\int {\mathcal D}A 
\exp \Big\{ i \big[ {\mathcal S}_{\rm em}(A) \big]\Big\}}
\;.
\end{equation}
The presence of the mirror may be introduced in more than one way; in the
previous definition of the functional averages, since the action is the one
for free space, we have implicitly assumed that it is dealt
with by a proper definition of the integration measure. Namely, the
integral is over fields satisfying perfect boundary conditions
on the mirror. Our choice of coordinates is such that the mirror occupies
the $x_3 = 0$ plane (for the sake of simplicity, we shall make no
distinction between lower and upper indices, from now on. Thus, \mbox{$x_3 \equiv
x^3 \equiv z$}).  From (\ref{Gammaam}), we see that we just
need to perform functional averages for {\em pairs\/} of fields (each
factor involves derivatives of the gauge field). Therefore, we shall only
need the gauge field propagator with perfect conductor boundary conditions
on the mirror.

Just before inserting the explicit expressions for the  EM field
correlators, it is convenient to perform an expansion in powers of the
departures about the average position of the atom, ${\mathbf r}_0$.  To that end, we set \mbox{${\mathbf
r}(t) \;=\; {\mathbf r}_0 + {\mathbf y}(t)$}, expand up to the second order
in ${\mathbf y}(t)$, discard terms which, by their very structure, cannot
contribute to the imaginary part of the effective action. 

Thus, with this in mind, we may present the expression for the effective
action, expanded to the second order in ${\mathbf y}(t)$, as follows:

\begin{equation}\label{eq:gammadec}
\Gamma \; =\; \Gamma_{EE} \,+\,\Gamma_{EB} \,+\, \Gamma_{BB} \;,
\end{equation}
where
\begin{align}\label{eq:gee1}
\Gamma_{EE} &=\, \frac{i e^2}{2 m} \;
	\int_{t,t'} \;   y_i(t) y_j(t') \; \Delta_\Omega(t-t') \nonumber \\ &\times \Big(
\frac{\partial^2}{\partial r_i \partial {r'}_j} 
\langle {\mathbf E}(t,{\mathbf r}) \cdot {\mathbf E}(t',{\mathbf r}')
\rangle \Big)\Big|_{{\mathbf r} =  {\mathbf r'} ={\mathbf r}_0}  \;,
\\
\Gamma_{EB} &=\, \frac{i e^2}{m} \;
	\int_{t,t'} \;   y_i(t) \dot{y}_j(t') \; \Delta_\Omega(t-t') \,
 \varepsilon_{kjl} \, \nonumber \\ & \times 
 \Big(\frac{\partial}{\partial r_i}  \langle  E_k(t,{\mathbf r}) \, B_l(t',{\mathbf r}')
\rangle \Big)\Big|_{{\mathbf r} =  {\mathbf r'} ={\mathbf r}_0} \;, \\ 
\Gamma_{BB} &=\, \frac{i e^2}{2 m} \;
	\int_{t,t'} \;   \dot{y}_i(t) \dot{y}_j(t') \; \Delta_\Omega(t-t') \,
\varepsilon_{kil} \varepsilon_{kjm} \nonumber \\ & \times 
\langle  B_l(t,{\mathbf r}) \, B_m(t',{\mathbf r}')
\rangle\Big|_{{\mathbf r} =  {\mathbf r'} ={\mathbf r}_0}  \;.
\end{align}
The previous formulae, valid in free space, hold true when a mirror is
present, the difference between those two situations being the form of the
EM field correlation functions. 
Let us first consider the free space case.

\subsection{Free space}

We evaluate each one of the three terms into which we have decomposed
$\Gamma$ in (\ref{eq:gammadec}) in turn. They involve different correlation
functions between components of the EM field in free space. Note that any
time-local term appearing in those correlation functions (namely, a
polynomial in $\delta(t-t')$ and its derivatives) will not contribute to
the imaginary part and we shall therefore discard them.  We shall use a
$(0)$ to denote the free space version of an object, to distinguish it from
the one when the perfectly conducting plane is present.


For $\Gamma_{EE}^{(0)}$  we have:
\begin{eqnarray}\label{eq:gee}
	\langle E_i(t,{\mathbf r}(t))  E_j(t',{\mathbf r}(t')) \rangle^{(0)}
	&=& \int \frac{d^3{\mathbf k}}{(2\pi)^3} 
	e^{i {\mathbf k} \cdot ({\mathbf r} - {\mathbf r'})}  \\
&\times &	k^2  (\delta_{ij} - \frac{k_i k_j}{k^2}) \Delta_k(t - t') \nonumber ,
\end{eqnarray}
where $\langle \ldots \rangle^{(0)}$ denotes correlation functions in free
space.
Here, $\Delta_k$ is defined as in (\ref{eq:defdelta}), with 
$k \equiv |{\mathbf k}|$ playing the role of $\Omega$: 
$\Delta_k(t-t') \equiv [\Delta_\Omega(t-t')]_{\Omega \to k}$. 
Therefore, from (\ref{eq:gee}) we derive:
\begin{align}
\Big(\frac{\partial^2}{\partial r_i \partial r'_j} 
 \langle {\mathbf E}(t,{\mathbf r}) & \cdot {\mathbf E}(t',{\mathbf r}')
\rangle^{(0)} \Big)\Big|_{{\mathbf r} =  {\mathbf r'} ={\mathbf r}_0} \nonumber \\
 & = 2 \int \frac{d^3{\mathbf k}}{(2\pi)^3} \,{\mathbf k}^2
\, k_i k_j \, \Delta_k(t - t')  \nonumber \\
& =\; \frac{2}{3} \, \int \frac{d^3{\mathbf k}}{(2\pi)^3}\, k^4
\, \delta_{ij} \, \Delta_k(t - t') \;.
\end{align}

Inserting this into $\Gamma_{EE}^{(0)}$, we note that the resulting expression will
contain the product $\Delta_\Omega \Delta_k$. This product may be simplified by using the property, valid for any pair $\Omega_1$, $\Omega_2$:
\begin{equation}
\Delta_{\Omega_1}(t-t') \, \Delta_{\Omega_2}(t-t') \;=\;
\frac{\Omega_1 + \Omega_2}{2 \, \Omega_1 \Omega_2} \, 
\Delta_{\Omega_1+\Omega_2}(t-t') \;.
\end{equation}
In our case, this leads to:
\begin{equation}
\Gamma_{EE}^{(0)} = \frac{i e^2}{6 m \Omega} \int_{t,t'} 
y_i(t) y_i(t')\int \frac{d^3{\mathbf k}}{(2\pi)^3} 
(\Omega + k) k^3 \Delta_{\Omega + k}(t-t')
\end{equation}
Writing then $\Delta_{\Omega + k}$ in terms of its Fourier transform, and Fourier transforming the departures, we obtain
\begin{equation}
\Gamma_{EE}^{(0)} = \frac{i e^2}{6 m \Omega} \int \frac{d\nu}{2\pi} 
\tilde{y}^*_i(\nu)\tilde{y}_i(\nu) \int \frac{d^3{\mathbf k}}{(2\pi)^3} 
(\Omega + k)  k^3  \widetilde{\Delta}_{\Omega + k}(\nu).
\end{equation}
The imaginary part of $\Gamma_{EE}$ is then straightforwardly obtained from the one of $\widetilde{\Delta}_{\Omega + k}$:
\begin{eqnarray}
{\rm Im}\big[\Gamma_{EE}^{(0)}\big] =  \frac{\pi e^2}{6 m \Omega}  & \int \frac{d\nu}{2\pi}
\tilde{y}^*_i(\nu)\tilde{y}_i(\nu) & \, \int \frac{d^3{\mathbf k}}{(2\pi)^3}  
(\Omega + k) \\ 
&& \times k^3  \delta\big[\nu^2 - (\Omega + k)^2 \big]\nonumber .
\end{eqnarray}
Performing the integration over ${\mathbf k}$,
\begin{eqnarray}
{\rm Im}\big[\Gamma_{EE}^{(0)}\big] &=& \int \frac{d\nu}{2\pi}\;  m_{EE}^{(0)}(\nu) |\tilde{\mathbf y}(\nu)|^2 \\
&=&\frac{e^2}{24 \pi m \Omega} \, \int \frac{d\nu}{2\pi}\; \theta(|\nu| - \Omega) \,
(|\nu| - \Omega)^5 \; 
|\tilde{\mathbf y}(\nu)|^2. \nonumber 
\end{eqnarray}


For the computation of  $\Gamma_{EB}^{(0)}$ we start from the correlation function:
\begin{eqnarray}
	\langle E_j(t,{\mathbf r})  B_l(t',{\mathbf r}') \rangle^{(0)}
	& = & i \epsilon_{jlm} \int  \frac{d^3{\mathbf k}}{(2\pi)^3} 
	e^{i {\mathbf k} \cdot ({\mathbf r} - {\mathbf r'})} \\
	&& \times k_m 
	\partial_t \Delta_k(t - t') \nonumber .
\end{eqnarray}
Upon insertion of this into the expression for $\Gamma_{EB}^{(0)}$, the product 
$\Delta_\Omega \partial_t \Delta_k$ arises. For this object we use the property:
\begin{equation}
\Delta_{\Omega_1}(t-t') \, \partial_t\Delta_{\Omega_2}(t-t') \;=\;
\frac{1}{2 \Omega_1} \, \partial_t\Delta_{\Omega_1 +\Omega_2}(t-t') \;,
\end{equation}
to get:
\begin{equation}
\Gamma_{EB}^{(0)} \;=\; \frac{i e^2}{3 m \Omega} \, \int_{t,t'} 
\, y_i(t) \dot{y}_i(t') \, \int \frac{d^3{\mathbf k}}{(2\pi)^3} \; 
k^2\; \partial_t \Delta_{\Omega + k}(t-t') \;.
\end{equation}
Integrating by parts and Fourier transforming,
\begin{equation}
\Gamma_{EB}^{(0)} = - \frac{i e^2}{3 m \Omega}  
\int \frac{d\nu}{2\pi} \tilde{y}^*_i(\nu)\tilde{y}_i(\nu)  \nu^2 
\int \frac{d^3{\mathbf k}}{(2\pi)^3}  k^2 
\widetilde{\Delta}_{\Omega + k}(\nu),
\end{equation}
whence the imaginary part then becomes:
\begin{eqnarray}
&{\rm Im}&\big[\Gamma_{EB}^{(0)}\big] = \int \frac{d\nu}{2\pi} |\tilde{\mathbf y}(\nu)|^2 m_{EB}^{(0)}(\nu)  \\
& = & -\frac{e^2}{12 \pi m \Omega} \int \frac{d\nu}{2\pi}
|\tilde{\mathbf y}(\nu)|^2  \theta(|\nu| - \Omega) 
|\nu| (|\nu| - \Omega)^4 \nonumber .
\end{eqnarray}


Finally, to evaluate $\Gamma_{BB}^{(0)}$, we need the correlator of two magnetic fields. It is straightforward to see that:
\begin{eqnarray}
\varepsilon_{kil} \varepsilon_{kjm}
\langle  B_l(t,{\mathbf r}) \, B_m(t',{\mathbf r}') \rangle^{(0)}
& = & \frac{4}{3} \, \delta_{ij} \,  \int \frac{d^3{\mathbf k}}{(2\pi)^3} 
	e^{i {\mathbf k} \cdot ({\mathbf r} - {\mathbf r'})} \nonumber \\  && \times k^2\,  
	\Delta_k(t - t') .
\end{eqnarray}
Therefore,
\begin{equation}
\Gamma_{BB}^{(0)} = \frac{2 i e^2}{3 m} 
	\int_{t,t'}   \dot{y}_i(t) \dot{y}_i(t')  \Delta_\Omega(t-t') 
 \int \frac{d^3{\mathbf k}}{(2\pi)^3}  k^2 
\Delta_k(t - t').
\end{equation}
Proceeding in an analogous fashion as for the previous two terms,
we find:
\begin{eqnarray}
{\rm Im}\big[\Gamma_{BB}^{(0)}\big] &=&  \int \frac{d\nu}{2\pi}\; 
|\tilde{\mathbf y}(\nu)|^2 \; m_{BB}^{(0)}(\nu) \\
&=& \frac{e^2}{12 \pi m \Omega} \int \frac{d\nu}{2\pi}
|\tilde{\mathbf y}(\nu)|^2 \; \theta(|\nu| - \Omega) 
\nu^2 (|\nu| - \Omega)^3.\nonumber 
\end{eqnarray}

Adding the three contributions to the imaginary part of the effective action, we get:

\begin{align}\label{eq:imgammafree}
{\rm Im}[\Gamma^{(0)}] =\; {\rm Im}[\Gamma_{EE}^{(0)}] \,+\,{\rm Im}[\Gamma_{EB}^{(0)}] 
\,+\, {\rm Im}[\Gamma_{BB}^{(0)}] \nonumber\\
=\;\frac{e^2}{24\pi m \Omega}\, \int_{-\infty}^{+\infty} \frac{d\nu}{2\pi}\; 
 |\tilde{\mathbf y}(\nu)|^2 \; \theta(|\nu| - \Omega) \nonumber \\ \times (|\nu| - \Omega)^3
\, (\nu^2 + \Omega^2).
\end{align}
This coincides with the result obtained in Ref.\cite{NetoMic}, if one
performs the angular integration of the probability distribution obtained there. 

\subsection{Perfect mirror}\label{ssec:perf}

We now evaluate the different terms contributing to the imaginary part of
the effective action when a perfect mirror is present. The difference is in
the form of the EM field correlation functions. They may be obtained from
the ones of the gauge field, which in turn can be constructed, for example,
by using the method of images. To that end, it is convenient to introduce
first a special notation to distinguish among spacetime coordinates.  We
shall use: $x\,=\,(x_\shortparallel , x_3)$, with $x_\shortparallel$
denoting \mbox{$x_0, \,x_1,\, x_2$}, the coordinates for which there is
translation invariance. When using indices, the ones from the beginning of
the Greek alphabet: $\alpha, \,\beta, \ldots$,  will be implicitly assumed
to run over the values $0$, $1$ and $2$. Besides, $a,\,b,\,\ldots$ will
take the values $1$ and $2$ (these appear when dealing with spatial
coordinates on the plane). 

Then, the correlator in the presence of the mirror,
\begin{equation}
\langle A_\mu(x) A_\nu(y) \rangle \;\equiv\; D_{\mu\nu}(x_\shortparallel -
	y_\shortparallel; x_3, y_3) \;,  
\end{equation}
may be written as follows:
\begin{eqnarray}
D_{\mu\nu}(x_\shortparallel - y_\parallel; x_3 , y_3) &= &  
D^{(0)}_{\mu\nu}(x_\shortparallel - y_\shortparallel; x_3 , y_3) \nonumber \\ & + &
D^{(R)}_{\mu\nu}(x_\shortparallel - y_\shortparallel; x_3 , y_3) \;,
\end{eqnarray}
where $D^{(0)}_{\mu\nu}$ is the gauge-field propagator in free space, and
$D^{(R)}_{\mu\nu}$ is the reflected contribution:
\begin{equation}
D^{(R)}_{\mu\nu} \;=\; - \, g_\mu^\alpha \, g_\nu^\beta \;
D^{(0)}_{\alpha\beta}(x_\shortparallel - y_\shortparallel; x_3 , - y_3) \;.
\end{equation}
Because of the fact that the EM field correlators will be the sum of two terms, the
first one identical to the free space one, and the second a ``reflection''
($R$) term, also the effective action and its imaginary part will share this property. 
Namely, 
\begin{align}
\Gamma & =\; \Gamma^{(0)} \,+\,\Gamma^{(R)} \;,\nonumber\\
\Gamma^{(R)} & =\; \Gamma_{EE}^{(R)} \,+\, \Gamma_{EB}^{(R)}  
\,+\, \Gamma_{BB}^{(R)} \;.
\end{align}

We now evaluate each one of the three reflection terms above, having in
mind that they are to be added to the free space terms; namely, they have
no meaning by themselves and in particular their imaginary parts could be
negative.


$\Gamma_{EE}^{(R)}$  is obtained by using the reflection term instead of the free
space correlator in the analogous formula we have already used for
free space. Indeed,
\begin{eqnarray}
\Gamma_{EE}^{(R)} \,=\, \frac{i e^2}{2 m} \;
	& \int_{t,t'} \;   & y_i(t)  y_j(t') \; \Delta_\Omega(t-t')  \\ &\times & \Big(
\frac{\partial^2}{\partial r_i \partial {r'}_j} 
\langle {\mathbf E}(t,{\mathbf r}) \cdot {\mathbf E}(t',{\mathbf r}')
\rangle^{(R)}\Big)\Big|_{{\mathbf r} =  {\mathbf r'} ={\mathbf r}_0}  \nonumber
\end{eqnarray}
Now, because of the presence of the mirror, three-dimensional rotation symmetry 
is lost. We will, as a consequence, have different contributions to 
$\Gamma_{EE}^{(R)}$ (and its imaginary part) depending on whether the
motion is parallel or normal to the plane. It is rather straightforward to
see that $\Gamma_{EE}^{(R)}$ becomes the sum of two
independent contributions, one for each kind of motion:
\begin{equation}
	\Gamma_{EE}^{(R)}[{\mathbf y}(t)]  \,=\,\Gamma_{EE,\shortparallel}^{(R)}[{\mathbf
	y_\shortparallel}(t)] \,+\,\Gamma_{EE,\perp}^{(R)}[y_3(t)]   \;,
\end{equation}
since:
\begin{align}
\Gamma_{EE}^{(R)}[{\mathbf y}(t)] \, & =\, \frac{i e^2}{2 m} \;
	\int_{t,t'} \; \Big\{ \frac{1}{2} \, 
	{\mathbf y}_\shortparallel(t) \cdot {\mathbf y}_\shortparallel(t') \; \Delta_\Omega(t-t') \\ & \times  \Big(
\frac{\partial^2}{\partial r_a \partial {r'}_a} 
\langle {\mathbf E}(t,{\mathbf r}) \cdot {\mathbf E}(t',{\mathbf r}')
\rangle^{(R)}\Big)\Big|_{{\mathbf r} =  {\mathbf r'} ={\mathbf r}_0}  \nonumber\\
 & +  \,	y_3(t)  y_3(t') \; \Delta_\Omega(t-t') \nonumber \\ & \times \Big(
\frac{\partial^2}{\partial r_3 \partial r'_3} 
\langle {\mathbf E}(t,{\mathbf r}) \cdot {\mathbf E}(t',{\mathbf r}')
\rangle^{(R)}\Big)\Big|_{{\mathbf r} =  {\mathbf r'} ={\mathbf r}_0} \Big\} \; \nonumber .
\end{align}

In particular, for parallel motion, we find:
\begin{widetext}
\begin{equation}
\Gamma_{EE,\shortparallel}^{(R)}[{\mathbf y_\shortparallel}(t)] =
	- \frac{i e^2}{4 m} \int_{t,t'} {\mathbf y_\shortparallel}(t)
	{\mathbf y_\shortparallel}(t') \int \frac{d^3{\mathbf
	k}}{(2\pi)^3}  {\rm cos}(2 k_3 a)  {\mathbf k_\shortparallel}^2
({\mathbf k_\shortparallel}^2 + 3 k_3^2) \Delta_\Omega(t-t')
 \Delta_k(t-t') \,,
\end{equation}
\end{widetext}
while for motion along the perpendicular, $x_3$ direction:
\begin{eqnarray}
\Gamma_{EE,\perp}^{(R)}[ y_3(t)] & = & 
	 \frac{i e^2}{2 m} \int_{t,t'}  y_3(t) y_3(t')  \int  \frac{d^3{\mathbf
	k}}{(2\pi)^3}  {\rm cos}(2 k_3 a)   k_3^2 \nonumber \\
&\times & ({\mathbf k_\shortparallel}^2 + 3 k_3^2) \Delta_\Omega(t-t')
 \Delta_k(t-t') \,.
\end{eqnarray}

By using an entirely analogous procedure to the one for the free space
part, we find the respective imaginary parts. Note that both can, and will,
depend on $a$, the distance of ${\mathbf r}_0$ to the mirror:

\begin{equation} 
{\rm Im}\big[\Gamma_{EE,\shortparallel}^{(R)}\big] = \int \frac{d\nu}{2\pi}\, |\tilde{\mathbf y}_\shortparallel(\nu)|^2 \;
m_{EE}^\parallel (\nu), 
\end{equation}
where
\begin{equation}
m_{EE}^\parallel (\nu) =   - \frac{e^2}{32 \pi
m \Omega}\,
	\theta(|\nu| - \Omega) (|\nu| - \Omega)^5  \,f_1[ (|\nu| - \Omega) a],
\end{equation}
with:
\begin{equation}
	f_1(x) \,=\; 3(\frac{1}{x^4} - \frac{1}{2 x^2}) {\rm cos}(2 x) \,+\, \frac{1}{2} (- \frac{3}{x^5} + \frac{11}{2 x^3}) {\rm sin}(2
	x) \,.
\end{equation}
For the perpendicular case, we have 

\begin{equation}
{\rm Im}\big[\Gamma_{EE,\perp}^{(R)}\big]  =   \int \frac{d\nu}{2\pi}\,  |\tilde{y}_3(\nu)|^2 \; m_{EE}^\perp(\nu), 
\end{equation}
where
\begin{equation}
m_{EE}^\perp(\nu) = \frac{e^2}{16 \pi
m \Omega}\, \int \frac{d\nu}{2\pi}\, 
	\theta(|\nu| - \Omega) (|\nu| - \Omega)^5  \,f_2[ (|\nu| - \Omega) a],
\end{equation}
with:
\begin{equation}
	f_2(x) \,=\; (- \frac{3}{x^4} + \frac{5}{2 x^2}) {\rm cos}(2 x) \,+\, 
	\frac{1}{4} (\frac{6}{x^5} - \frac{13}{x^3}+ \frac{6}{x}) {\rm sin}(2
	x) \,.
\end{equation}


The term which involves the mixed correlator between the electric and magnetic fields, may be written as follows
\begin{eqnarray}
\Gamma_{EB}^{(R)} & = & \frac{i e^2}{m} \;
	\int_{t,t'} \; \int
	\frac{d^3{\mathbf k}}{(2\pi)^3}  e^{2 i k_3 a} \,
	\Delta_\Omega(t-t') \partial_t\Delta_k(t-t')   \nonumber \\ 
& \times &	y_j(t) \dot{y}_l(t') \; 
(k_j\,\delta_{l a} k_a - 2 k_j \delta_{l 3} k_3 )  .
\end{eqnarray}
The imaginary parts for parallel and normal motion become

\begin{equation}
{\rm Im}\big[\Gamma_{EB,\shortparallel}^{(R)}\big]  =  \int \frac{d\nu}{2\pi}\, 
	|\tilde{\mathbf y}_\shortparallel(\nu)|^2 \; m_{EB}^\parallel (\nu), 
	\end{equation}
where
\begin{equation}
m_{EB}^\parallel (\nu) =  - \frac{e^2}{8 \pi
m \Omega} \theta(|\nu| - \Omega) \,
	|\nu| (|\nu| - \Omega)^4  \,f_3[ (|\nu| - \Omega) a]\;,
\end{equation}
with:
\begin{equation}
	f_3(x) \,=\;  - \frac{1}{4 x^2} {\rm cos}(2 x) \,+\, 
	\frac{1}{8 x^3} \, {\rm sin}(2 x) \,,
\end{equation}
and

\begin{equation} 
{\rm Im}\big[\Gamma_{EB,\perp}^{(R)}\big]   =  \int \frac{d\nu}{2\pi}\, 
	|\tilde{y}_3(\nu)|^2 \; m_{EB}^\perp (\nu), 
	\end{equation}
where, 

\begin{equation}
  m_{EB}^\perp (\nu) =  \frac{e^2}{8\pi
m \Omega}\,  \theta(|\nu| - \Omega) \,|\nu| (|\nu| - \Omega)^4  \,f_4[ (|\nu| - \Omega) a], 
\end{equation}
with:
\begin{equation}
	f_4(x) \,=\; \frac{1}{x^2} \,{\rm cos}(2 x) \,-\, 
	(\frac{1}{2 x^3} -\frac{1}{x}) {\rm sin}(2 x) \,.
\end{equation}


Finally, for  {$\Gamma_{BB}^{(R)}$ we have: 

\begin{align}
\Gamma_{BB}^{(R)} & =\, - \frac{i e^2}{2 m} \;
	\int_{t,t'} \; \int
	\frac{d^3{\mathbf k}}{(2\pi)^3}  e^{2 i k_3 a} \,
	\Delta_\Omega(t-t') \Delta_k(t-t') \nonumber \\ & \times \Big\{  \, 
	\dot{y}_a(t) \dot{y}_a(t') \; 
({\mathbf k_\shortparallel}^2 - k_3^2)
  -  2 \,k_3^2 \,\dot{y}_3(t)  \dot{y}_3(t') \; \Big\} \;.
\end{align}
and the respective imaginary parts for parallel and normal motion,

\begin{equation} 
{\rm Im}\big[\Gamma_{BB,\shortparallel}^{(R)}\big] =   \int \frac{d\nu}{2\pi}\, 
	|\tilde{\mathbf y}_\shortparallel(\nu)|^2 m_{BB}^\parallel (\nu), 
	\end{equation}
where
\begin{equation}
 m_{BB}^\parallel (\nu) = - \frac{e^2}{16 \pi
m \Omega} \theta(|\nu| - \Omega) 
	\nu^2 (|\nu| - \Omega)^3  \,f_5[ (|\nu| - \Omega) a],
\end{equation}
with:
\begin{equation}
	f_5(x) \,=\;  - \frac{1}{x^2} {\rm cos}(2 x) \,+\, \frac{1}{2}
	 (\frac{1}{x^3} - \frac{1}{x}) {\rm sin}(2 x) \,,
\end{equation}
and
\begin{equation}
{\rm Im}\big[\Gamma_{BB,\perp}^{(R)}\big]  = \int \frac{d\nu}{2\pi}\, 
	|\tilde{y}_3(\nu)|^2 \; m_{BB}^\perp (\nu),
	\end{equation}
	where
\begin{equation}
m_{BB}^\perp (\nu) =  \frac{e^2}{16 \pi
m \Omega}  \theta(|\nu| - \Omega) \,\nu^2 	(|\nu| - \Omega)^3  \,f_6[ (|\nu| - \Omega) a]\;,
\end{equation}
with:
\begin{equation}
	f_6(x) \,=\; \frac{1}{x^2} \,{\rm cos}(2 x) \,+\, 
	(-\frac{1}{2 x^3} +\frac{1}{x}) {\rm sin}(2 x) \,.
\end{equation}

In Fig. \ref{figEE} we plot $m_1 = 1 +  m_{EE}^\parallel/m_{EE}^{(0)}$ for the parallel motion and $m_2=  1 + m_{EE}^\perp/m_{EE}^{(0)}$ for the 
normal one, both as functions of the dimensionless variable $x = a (\vert \nu\vert - \Omega)$. In both cases, the contribution from $m_{EE}^{\parallel,\perp}$ goes to zero as $a \vert \nu\vert \rightarrow \infty$. 
In the other limit, i.e., when $x \rightarrow 0$, we get $m_{EE}^\parallel/m_{EE}^{(0)} = - 7/10$ and $m_{EE}^\perp/m_{EE}^{(0)} = 11/10$. This case 
appears as qualitatively similar to the Dirichlet contribution reported in Ref. \cite{PaperI}, (see comment in \cite{footnote}).  Figs. \ref{figEB} and \ref{figBB} show similar behaviour than the previous one. 
The rates goes to zero in the large limit and $m_{EB}^\parallel/m_{EB}^{(0)} = 1/2$, $m_{EB}^\perp/m_{EB}^{(0)} = -1$, 
and $m_{BB}^\parallel/m_{BB}^{(0)} = -1/4$ and $m_{BB}^\perp/m_{BB}^{(0)} = 1/2$ when $x\rightarrow 0$.  These limits show a different behaviour with respect to the Dirichlet and Neumann cases
reported in \cite{PaperI}. 

\begin{figure}
\begin{center}
\includegraphics[scale=0.5]{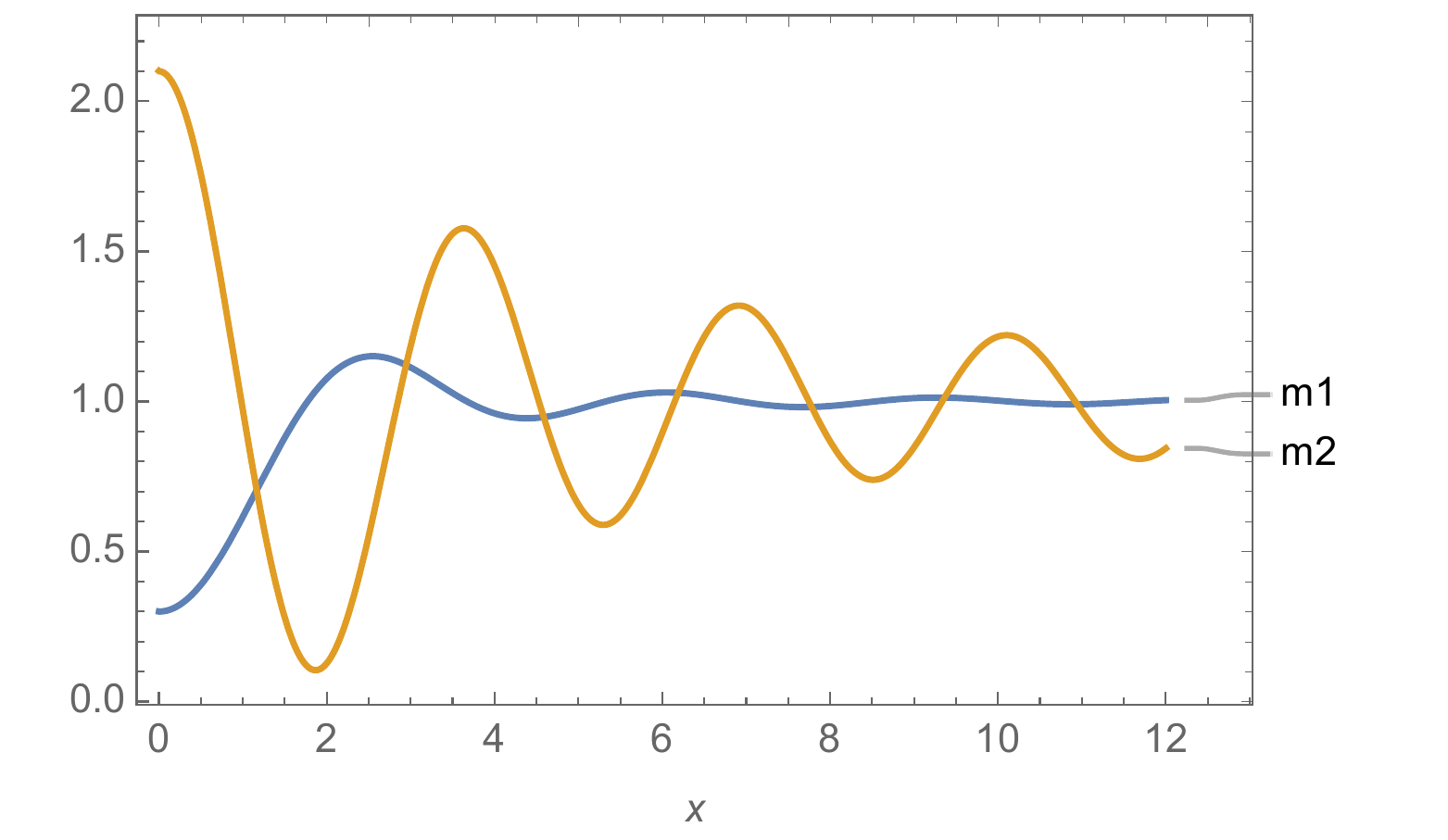}
\caption {We plot  $m_1 = 1 +  m_{EE}^\parallel/m_{EE}^{(0)}$ and  $m_2=  1 + m_{EE}^\perp/m_{EE}^{(0)}$  as a function of the dimensionless $x = a (\vert \nu\vert - \Omega)$.}
\label{figEE}
\end{center}
\end{figure}

\begin{figure}
\begin{center}
\includegraphics[scale=0.5]{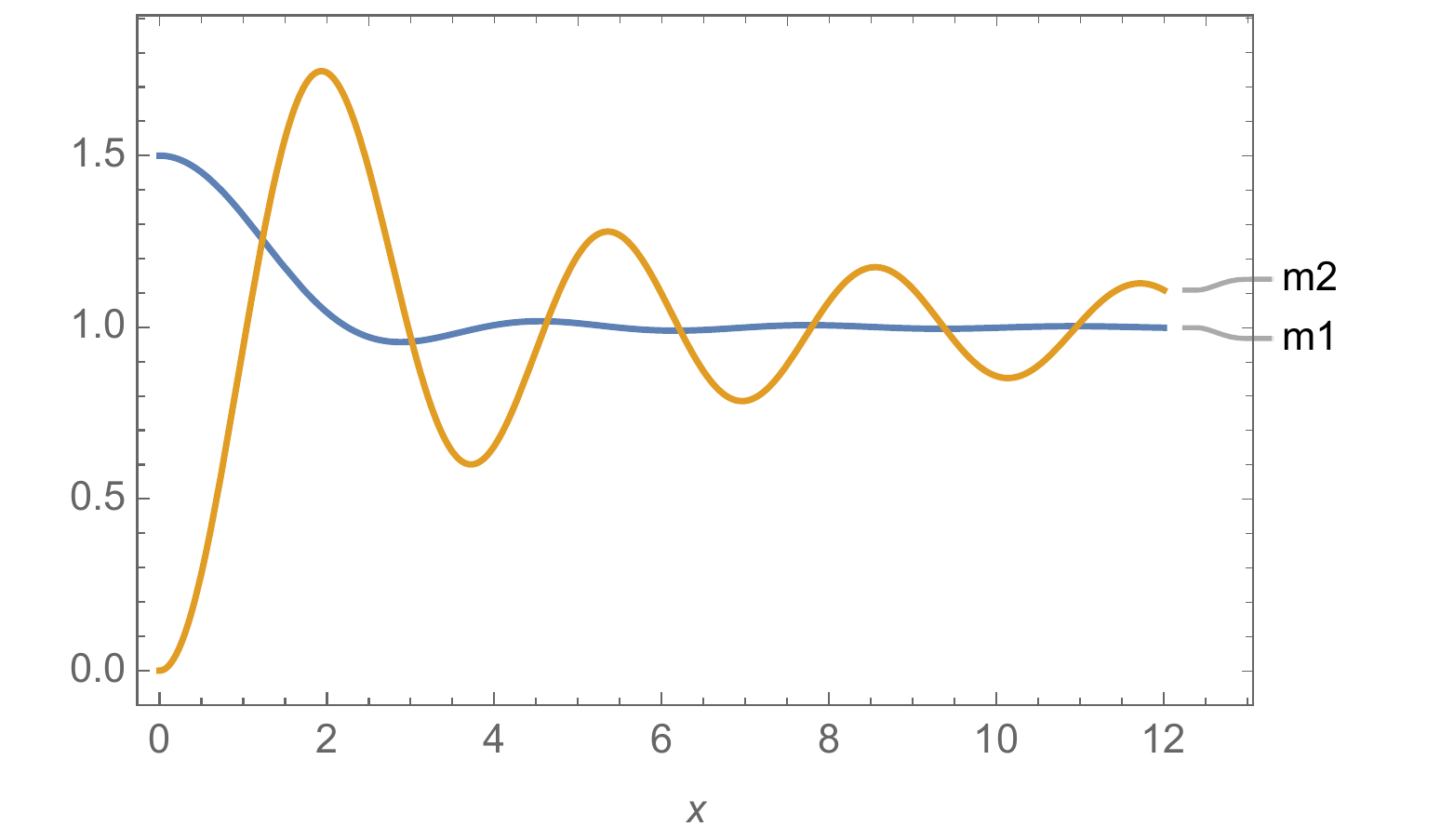}
\caption {$m_1 = 1 + m_{EB}^\parallel/m_{EB}^{(0)}$ and  $m_2=  1 + m_{EB}^\perp/m_{EB}^{(0)}$ as a function of the dimensionless $x = a (\vert \nu\vert - \Omega)$.}
\label{figEB}
\end{center}
\end{figure}

\begin{figure}
\begin{center}
\includegraphics[scale=0.5]{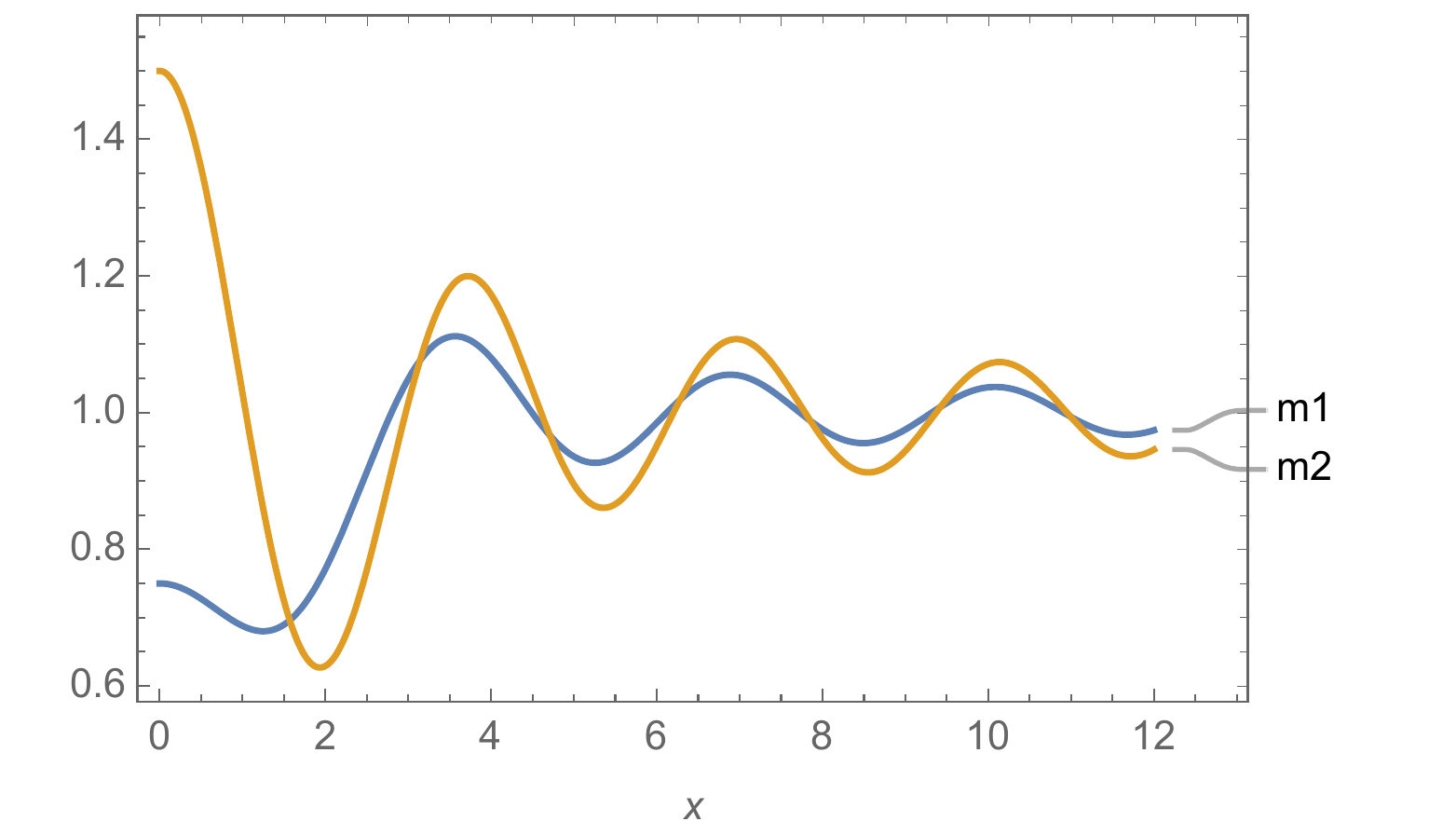}
\caption {Ratios  $m_1 = 1 + m_{BB}^\parallel/m_{BB}^{(0)}$ and  $m_2=  1 + m_{BB}^\perp/m_{BB}^{(0)}$ as a function of the dimensionless $x = a (\vert \nu\vert - \Omega)$.}
\label{figBB}
\end{center}
\end{figure}


\section{Transition amplitudes}\label{sec:amplitudes}

Let us now study the transition amplitudes and probabilities for the EM field model. The first-order transition matrix will now be given by:
\begin{equation}
	T_{fi} \;\equiv\;  e \, \int dt \, \langle f| 
	{\mathbf x}(t) \cdot [ {\mathbf E}(t,{\mathbf r}(t)) +\dot {\mathbf r}(t) \times {\mathbf B}(t,{\mathbf r}(t))] | i \rangle \;,
\end{equation}
where 
\begin{equation}
\vert i\rangle= \vert i_a\rangle \otimes \vert i_{EM}\rangle \quad\quad \vert f\rangle= \vert f_a\rangle \otimes \vert f_{EM}\rangle\,\, ,
\end{equation}
with the "$a$" and "$EM$" indices denoting the atom and electromagnetic field states.

For the electron's degrees of freedom we have, in the interaction picture
\begin{equation}
	{\mathbf x}(t) \;=\;\frac{1}{\sqrt{2 m \, \Omega}} \left( 
	{\mathbf a} e^{-i \Omega t} \,+\, {\mathbf a}^\dagger e^{i \Omega t}\right) \;.
\end{equation}
Here, ${\mathbf a} = \sum_{l=1}^3 a_l \; \widehat{\mathbf e}_l$, where 
$\widehat{\mathbf e}_l$ are three orthonormal vectors, since the Hamiltonian for the electron is essentially a three-dimensional harmonic oscillator.  This implies that, when considering a transition from the vacuum to an excited state, that process will introduce a spatial direction, in other words, a polarization.
On the other hand, for the gauge field in the Coulomb gauge we use the expansion:
\begin{equation}
{\mathbf A}(x)= \int d^2{\mathbf k}_\shortparallel \int_0^\infty  dk_z  \sum_{\lambda}
\left[ {\alpha}_\lambda({\mathbf k}) {\mathbf f}^{(\lambda)}_{\mathbf k}(x) 
+ {\mbox h.c.} \right],
\end{equation}
where $\lambda$ sums over the two independent modes for each ${\mathbf k}$, which are consistent with the perfect conductor condition at $z=0$. In this gauge, that amounts to a vanishing, on that plane, of the components of the vector potential which are parallel to that surface.

Including a global factor to normalize the states, we may write those modes as follows
(see, for example \cite{modes}):
\begin{align}
{\mathbf f}^{(1)}_{\mathbf k}(x) &=\; e^{-i k t} \sqrt{\frac{2}{(2\pi)^3 k}} \;
\big(\widehat{\mathbf k}_\shortparallel \times \widehat{\mathbf z} \big)
\sin(k_z z) \, 
e^{i {\mathbf k}_\shortparallel \cdot {\mathbf x}_\shortparallel}
\nonumber\\
{\mathbf f}^{(2)}_{\mathbf k}(x) &=\; e^{-i k t} \sqrt{\frac{2}{(2\pi)^3 k }} \;
k^{-1} \,\big[ \widehat{\mathbf z} \; |{\mathbf k}_\shortparallel | \cos(k_z z) 
\nonumber \\ & -\,i \,\widehat{\mathbf k}_\shortparallel \, k_z \, \sin(k_z z) \big]
e^{i {\mathbf k}_\shortparallel \cdot {\mathbf x}_\shortparallel} \;,
\end{align}
($\widehat{\mathbf k}_\shortparallel$ and $\widehat{\mathbf z}$ denote unit vectors). The notation
\begin{equation}
{\mathbf f}^{(\lambda)}_{\mathbf k}(x) =\; N_k\, e^{-i k t}{\mathbf g}^{(\lambda)}_{\mathbf k}(x)\,\; ,\,\; N_k=\sqrt{\frac{2}{(2\pi)^3 k}}
\end{equation}
will be useful in what follows.

\subsection{The decay process}
Let us now consider a decay process, in which the initial state of the atom is an excited state  and the EM field is in vacuum
\begin{equation}
\vert i_a\rangle =a_l^\dagger \vert 0_a\rangle\quad\quad \, \vert i_{EM}\rangle=\vert 0_{EM}\rangle ,
\end{equation}
while final states are
\begin{equation}
\vert f_a\rangle =\vert 0_a\rangle\quad\quad \,   \vert f_{EM}\rangle={\alpha}^\dagger_\lambda({\mathbf k})|0_{EM}\rangle.
\end{equation}
 Note that for the electronic transition we have in principle three independent polarizations (not necessarily along the three coordinate axis), so we have to  choose a polarization for the excited state of the electron, and also for the final state of the  EM field.

The matrix elements for the decay process then read

\begin{widetext}
\begin{eqnarray}
T_{fi}^{\rm{(dec)}}({\mathbf k}, l,\lambda) \;&=&\; 
\frac{e}{\sqrt{2 m\, \Omega}} \; 
\int_{-\infty}^{+\infty} dt \, e^{-i t \Omega} \, 
\langle 0_{EM}|\alpha_\lambda({\mathbf k})\,
\widehat{\mathbf e}_l\,  \cdot[{\mathbf E}(t,{\mathbf r}(t))+\dot{\mathbf r}(t)\times  {\mathbf B}(t,{\mathbf r}(t))]|0_{EM}\rangle\nonumber\\
&=& \; \frac{e}{\sqrt{2 m\, \Omega}} \; 
\int_{-\infty}^{+\infty} dt \, e^{-i t \Omega} \,
\widehat{\mathbf e}_l\,  \cdot\left[-\partial_t{\mathbf f}_{\mathbf k}^{(\lambda)*}(x)+\dot{\mathbf r}(t)\times\left({\mathbf\nabla}\times{\mathbf f}_{\mathbf k}^{(\lambda)*}(x)\right)\right]_{x=(t,{\mathbf r}(t))}.
\end{eqnarray}
\end{widetext}

We now expand 
the results up to the second order in $\mathbf y(t)$, which is the departure from $\mathbf r_0 = (0,0,a)$. Note that, as the matrix elements have a contribution at zeroth order, it is necessary to expand them up to the second order, to compute consistently the decay probabilities beyond the static case. We denote the different orders by
\begin{equation}
T_{fi}^{\rm{(dec)}}=T_{fi}^{\rm{(dec,0)}}+T_{fi}^{\rm{(dec,1)}}+T_{fi}^{\rm{(dec,2)}}\, .
\end{equation}

Performing the expansion  we obtain
\begin{eqnarray}\label{Tfiorders}
& T_{fi}^{\rm{(dec,0)}}&({\mathbf k}, l,\lambda) =  \frac{-2 \pi i e N_k k}{\sqrt{2 m\, \Omega}} \delta(\Omega-k)
\widehat{\mathbf e}_l
 \cdot \mathbf g^{(\lambda)*}_{\mathbf k} (\mathbf x)\vert_{\mathbf x=\mathbf r_0}\nonumber\\
&T_{fi}^{\rm{(dec,1)}}&({\mathbf k},l,\lambda) =
\frac{e N_k }{\sqrt{2 m\, \Omega}} 
\widehat{\mathbf e}_l
 \cdot[(- i k)\widetilde y_j(k-\Omega)\partial_j\mathbf g^{(\lambda)*}_{\mathbf k} (\mathbf x) \nonumber\\
 &&-  i  (k-\Omega)\tilde{\mathbf y}(k-\Omega)\times
(\mathbf \nabla\times\mathbf g^{(\lambda)*}_{\mathbf k} (\mathbf x))]\vert_{\mathbf x=\mathbf r_0} \nonumber\\
& T_{fi}^{\rm{(dec,2)}}&({\mathbf k}, l,\lambda) =
  A_{fi}({\mathbf k}, l,\lambda) +  B_{fi}({\mathbf k}, l,\lambda)\, ,
\end{eqnarray}
with
\begin{eqnarray}
A_{fi}({\mathbf k}, l,\lambda)&=&
\frac{-i e N_k k }{2\sqrt{2 m\, \Omega}} \int_{\-\infty}^{\infty} dt
e^{-i t (\Omega-k)}y_i(t)y_j(t)\nonumber\\
&&\widehat{\mathbf e}_l
 \cdot\partial_i\partial_j\mathbf g^{(\lambda)*}_{\mathbf k} (\mathbf x)\vert_{\mathbf x=\mathbf r_0}\nonumber\\
B_{fi}({\mathbf k}, l,\lambda)&=&
\frac{(-1)^\lambda i e N_k k }{\sqrt{2 m\, \Omega}} \int_{\-\infty}^{\infty} dt
e^{-i t (\Omega-k)}y_j(t)\dot{\mathbf y}(t)\nonumber\\
&&\cdot (\widehat{\mathbf e}_l\times
\partial_j\mathbf g^{(\lambda ')*}_{\mathbf k} (\mathbf x_\parallel,z-\frac{\pi}{2 k_z})\vert_{\mathbf x=\mathbf r_0}\, .
\end{eqnarray}
Here we used the notation $\lambda'=1$ for $\lambda=2$ and vice versa.

In practice, we expect the experiments not to detect the polarization state of the excited state of the atom; therefore, it makes sense to consider the sum over the $3$ posible values of $l$ when evaluating the probabilities. Namely, we obtain results for the probabilities which depend only the polarization of the photon. Therefore, for an unpolarized initial state, the total decay probability reads
\begin{equation}\label{dPifT}
dP_{fi}^{\rm dec}(\mathbf k)=dP_{fi}^{\rm dec}(\mathbf k,1)+dP_{fi}^{\rm dec}(\mathbf k,2)
\end{equation}
with
\begin{equation}\label{dPiflambda}
dP_{fi}^{\rm dec}(\mathbf k,\lambda)=\frac{1}{3}d^3\mathbf k
\sum_{l=1}^{3}|T_{fi}^{\rm dec}(\mathbf k,l,\lambda)|^2\, .
\end{equation}

Note that, when computing $|T_{fi}^{\rm dec}(\mathbf k,l,\lambda)|^2$,
there will be a contribution of zeroth order that gives the emission
probability  for a static atom. By energy conservation, this probability is
proportional to $\delta(\Omega-k)$. It is corrected by the second order
contribution to the matrix element.  On the other hand, the first order
contribution to the transition amplitude produces an emission probability
that, for a center of mass oscillation with frequency $\Omega_{\rm cm}$, has
lateral peaks at $k=\Omega\pm \Omega_{\rm cm}$. This is the main qualitative
change induced by the center of mass motion on the spectrum of emitted
photons.

We now assume a normal motion for the center of mass of the atom, with $\tilde y_j=\tilde y_\perp\delta_{j3}$. 
The zeroth order contribution $T_{fi}^{(dec,0)}$ in Eq.\eqref{Tfiorders} generates the spontaneous emission probability for a static atom. It reads
\begin{equation}\label{static pro}
dP_{fi}^{(\rm dec,0)}(\mathbf k)=
\frac{T e^2 N_k^2}{2m\Omega}2\pi\delta(\Omega-k)k^2 \sum_\lambda
\vert{\mathbf g}^{(\lambda)}_{\mathbf k} (\mathbf r_0)\vert^2\, d^3\mathbf k\, .
\end{equation}

The second order contribution in Eq.\eqref{Tfiorders}, when multiplied by the zeroth order, produces a correction to the static probability in Eq.\eqref{static pro} that is given by
\begin{eqnarray}\label{static corr}
dP_{fi}^{(\rm sta,2)}(\mathbf k)&=&
-\frac{ e^2 N_k^2}{2m\Omega}2\pi\delta(\Omega-k)k^2 k_z^2\sum_\lambda
\vert{\mathbf g}^{(\lambda)}_{\mathbf k} (\mathbf r_0)\vert^2\nonumber\\ 
&&\times \left(\int_{-\infty}^\infty dt\, 
y_\perp^2(t)\right)\, d^3\mathbf k\, .
\end{eqnarray}
Note that the position of the peak is independent of the center of mass motion of the atom.

We now consider the novel contribution to the emission probability coming from  $T_{fi}^{(dec,1)}$ in Eq.\eqref{Tfiorders}. It is given by
\begin{widetext}
\begin{eqnarray}\label{Pdecperp1}
dP_{fi}^{\rm (dyn,2)}(\mathbf k,1) &=& \frac{e^2}{12\pi^2m\Omega}\, |\tilde y_\perp(k - \Omega)|^2\Omega^2 k^3
\cos^2\left(k a \cos\theta)\right)\cos^2\theta\sin\theta \,d\theta\, dk \nonumber \\
&\equiv & \frac{e^2}{12\pi^2m\Omega}\, |\tilde y_\perp(k - \Omega)|^2 k^3 \, p_1(ka, \Omega a, \theta)\, \sin\theta \,d\theta\, dk 
\end{eqnarray}
and
\begin{eqnarray}\label{Pdecperp2}
dP_{fi}^{\rm (dyn,2)}(\mathbf k,2)&=&\frac{e^2}{12\pi^2m\Omega}\, |\tilde y_\perp(k - \Omega)|^2 k^3[k^2\sin^2\theta\cos^2\theta\sin^2\left(
k a\cos\theta\right)\nonumber\\
&+& (\Omega-k\sin^2\theta)^2\cos^2\left(k a\cos\theta\right)]\sin\theta \,d\theta\, dk \nonumber \\
&\equiv & \frac{e^2}{12\pi^2m\Omega}\, |\tilde y_\perp(k - \Omega)|^2 k^3 \, p_2(ka, \Omega a, \theta)\, \sin\theta \,d\theta\, dk. 
\end{eqnarray}
\end{widetext}
In both equations we have used spherical coordinates in $\mathbf k$-space and integrated over the angle $\varphi$ (by symmetry, the results do not depend on this angle). As the correction in Eq.\eqref{static corr},  these are of course contributions quadratic in the amplitude of the center of mass motion. In Fig. \ref{probang} we plot the total contribution to the emission probability $p_1(ka, \Omega a, \theta) + p_2(ka, \Omega a, \theta)$ per unit of solid angle for two different values of $ka$ at a fixed value of $\Omega a = 10$. 

\begin{figure}
    \centering
     \begin{subfigure}[b]{0.2\textwidth}
        \centering
        \includegraphics[width=\linewidth,trim ={.2cm 0 .5cm .2cm}]{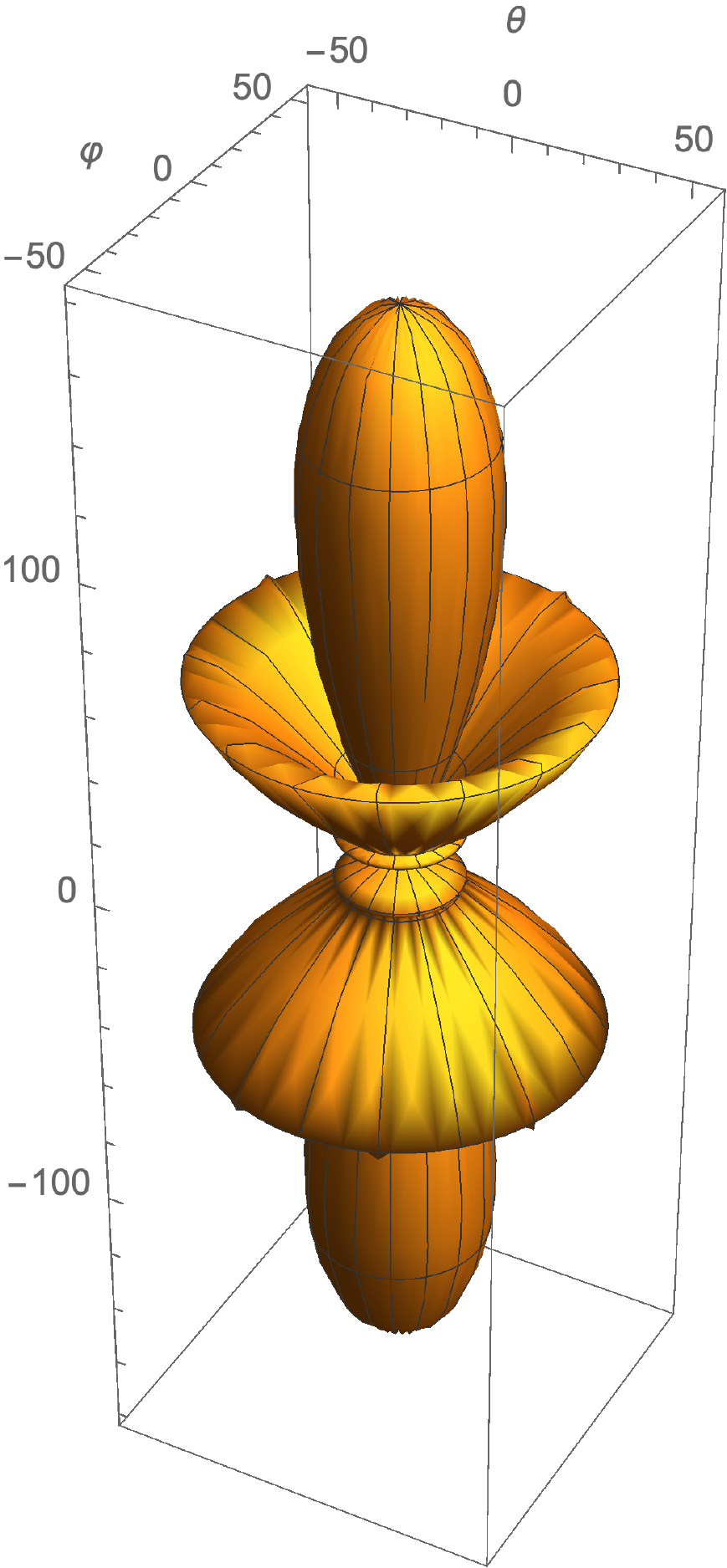}
        \caption{$ k a = 9, \Omega a = 10$}
    \end{subfigure}
    \hfill
    \begin{subfigure}[b]{0.25\textwidth}
        \centering
        \includegraphics[width=\linewidth,trim ={.2cm  0 .2cm .2cm}]{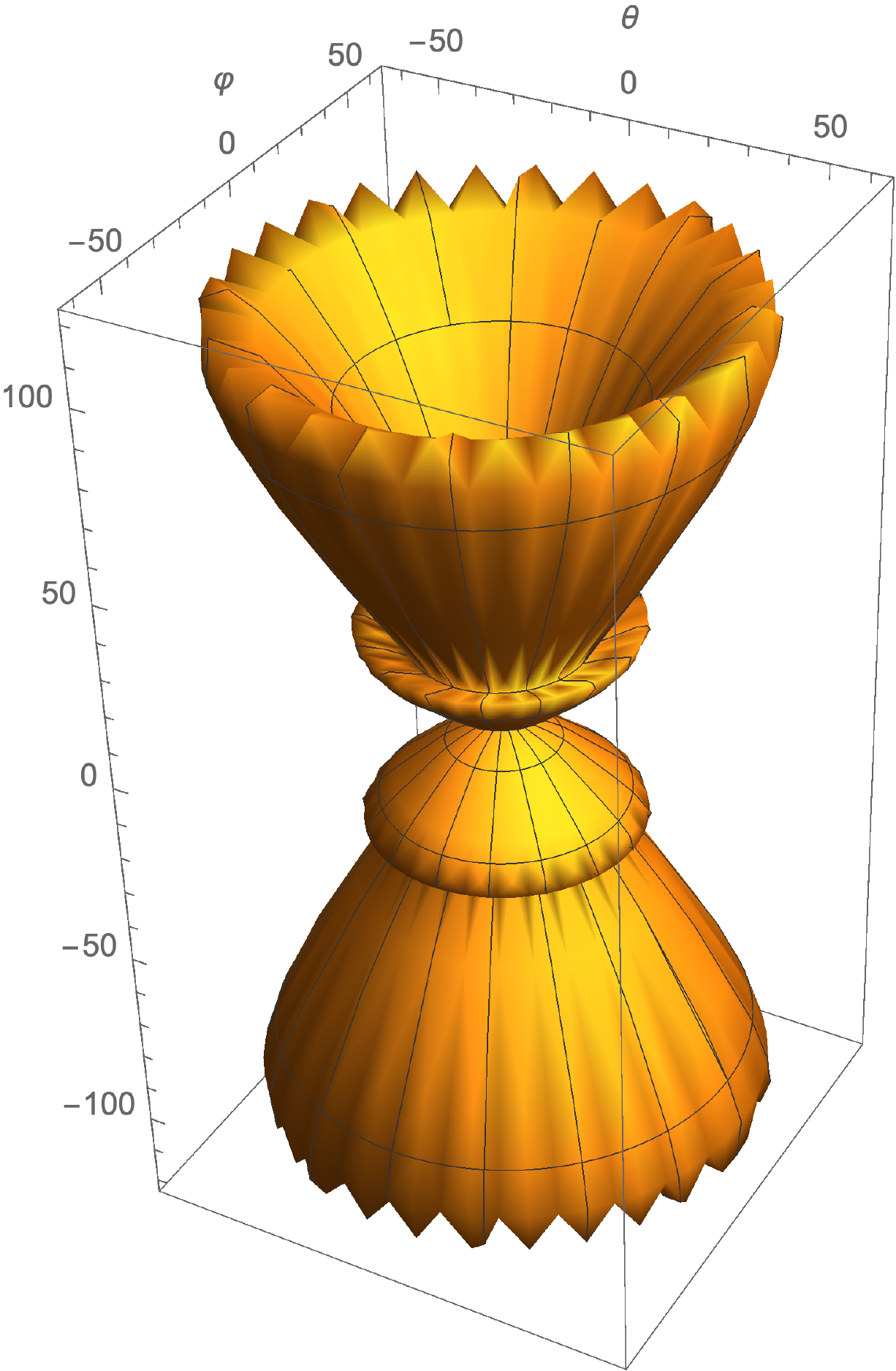}
        \caption{$ k a = 11, \Omega a = 10$}
\end{subfigure}
\caption{Total contribution to the emission probability $p_1(ka, \Omega a, \theta) + p_2(ka, \Omega a, \theta)$ per unit of solid angle as a function of spherical angle $\theta$. We show 
two different values of $ka$ at a fixed value of $\Omega a = 10$. }
\label{probang}
\end{figure}

The total decay probability can be obtained by summing Eqs.\eqref{Pdecperp1} and\eqref{Pdecperp2} and integrating the $\theta$ angle. The result is 

\begin{widetext}
\begin{eqnarray}
\frac{dP_{fi}^{\rm (dyn,2)}}{dk} & = & \frac{e^2}{144 \pi^2m\Omega a^5}\, |\tilde y_\perp(k - \Omega)|^2 \Big\{
8 a^5 k^3 \left(k^2-2 k \Omega +2 \Omega ^2\right)\\
&& +6 a k \left(a^2 (k+\Omega
   )^2-6\right) \cos (2 a k)+3 \left(4 a^4 k^2 \Omega ^2-a^2 \left(9 k^2+2 k
   \Omega +\Omega ^2\right)+6\right) \sin (2 a k)\Big\}\, .\nonumber
\end{eqnarray}
\end{widetext}

If we assume that the normal displacement $ y_\perp(t)$ is an oscillatory function $ y_\perp(t) =  y_\perp^0 \sin(\Omega_{\rm cm} t)$, where $\Omega_{\rm cm}$ is the frequency of the center of mass, the spectrum of the emitted photons has peaks at $\Omega, \Omega\pm\Omega_{\rm cm}$. In Fig.\ref{probang} we choose two different values of $ka$ ($ka = 9$ in Fig.\ref{probang}(a) and 
$ka = 11$ in Fig.\ref{probang}(b)) in order to show that just beyond adiabatic approximation ($ka \approx \Omega a$) the emission probability is non-symmetric with respect to the central emission peak.  


In a recent work \cite{Passante}, the decay probability of an atom in front
	of an oscillating mirror has been computed using the adiabatic
	approximation $\Omega \gg\Omega_{\rm cm}$. In this limit, our
	results for the moving atom have the same structure: the spectrum
	of the emitted photons has the above mentioned peaks, the decay
	probability decreases with the distance to the plane, and it shows
	oscillations with a frequency $2 k a\simeq 2 \Omega a$. There is a
	disagreement, however, between the coefficients of the terms
	appearing in our result and the ones of \cite{Passante}.  We have
	verified that this difference comes from the  R\"ontgen term.
	Indeed, omitting this contribution both results coincide.  For the
	case of a moving atom, it is well known that this interaction term
	is crucial for Lorentz covariance.  It would be interesting to
	check if it also appears for the case of a moving mirror and static
	atom, in the next to leading order of the adiabatic approximation.  Note that the boundary conditions for the electromagnetic field on a moving perfect mirror has a velocity-dependent term \cite{PMN96}.

The adiabatic approximation is justified when $\Omega \gg\Omega_{\rm cm}$. It
is noteworthy that, for some physical systems,  this inequality may be
violated: for Rydberg or artificial atoms may have $\Omega$  of the order
of GHz, and mechanical resonators may attain such frequencies. In this
situation, the decay probabilities may be qualitatively different from
those obtained in the adiabatic approximation.

%

\subsection{Excitation process}

We now consider the probability of excitation of an atom that is initially in its ground state.
 This excitation is accompanied by the emission of a photon. The initial states read
 \begin{equation}
\vert i_a\rangle =\vert 0_a\rangle\quad\quad \, \vert i_{EM}\rangle=\vert 0_{EM}\rangle ,
\end{equation}
while final states are
\begin{equation}
\vert f_a\rangle =a_l^\dagger\vert 0_a\rangle\quad\quad \,   \vert f_{EM}\rangle={\alpha}^\dagger_\lambda({\mathbf k})|0_{EM}\rangle.
\end{equation}

It is not necessary to repeat all calculations. Indeed, the matrix elements for the excitation process can be obtained from those of the decay process just changing the sign of the frequency 
$\Omega$, that takes into account the changes in the initial and final states of the atom. This change of the sign produces the expected threshold for the center of mass frequency, and the excitation occurs only above it. Therefore, the zeroth order in the transition amplitudes is absent for this process. 


From Eqs.\eqref{Pdecperp1} and \eqref{Pdecperp2} we obtain
\begin{widetext}
\begin{equation}\label{Pexperp1}
dP_{fi}^{\rm exc}(\mathbf k,1)=\frac{e^2}{12\pi^2m\Omega}\, |\tilde y_\perp(k + \Omega)|^2\Omega^2 k^3
\cos^2\left(k a \cos\theta)\right)\cos^2\theta\sin\theta \,d\theta\, dk \, 
\end{equation}
and
\begin{eqnarray}\label{Pexperp2}
dP_{fi}^{\rm exc}(\mathbf k,2)&=&\frac{e^2}{12\pi^2m\Omega}\, |\tilde y_\perp(k + \Omega)|^2 k^3[k^2\sin^2\theta\cos^2\theta\sin^2\left(
k a\cos\theta\right)\nonumber\\
&+& (\Omega +k\sin^2\theta)^2\cos^2\left(k a\cos\theta\right)]\sin\theta \,d\theta\, dk \, 
\end{eqnarray}
\end{widetext}
As before, in both equations we have used spherical coordinates in $\mathbf k$-space and integrated over the angle $\varphi$.


\section{Conclusions}\label{sec:conc}
In this paper we considered the interaction between an accelerated atom near a perfect mirror and the vacuum fluctuations of the electromagnetic field. We first computed the  vacuum persistence probability, and then the probabilities for excitation and decay,  for an atom that is initially in its ground or first excited state, respectively. The results generalize our previous work in which we studied, as a toy model, a quantum scalar field instead of the full electromagnetic field.

We have compared our results for an atom in perpendicular motion  with respect to the mirror, with those in which the atom is at rest and the mirror is oscillating. Up to the lowest order adiabatic approximation, the R\"ontgen current does not appear for a moving mirror \cite{Passante}, and this is a source of discrepancy between the results for both situations. It would be interesting to check whether the next to leading order adiabatic correction for the case of a moving mirror
restores the equivalence between these two different physical situations or not. 

Our results for the moving atom are valid beyond the adiabatic approximation, and we have pointed out that, for artificial or Rydberg atoms, this approximation may be violated. Therefore,  one could observe signs of non-adiabaticity in the spectrum of emitted particles

If the atom has a center of mass motion parallel to the mirror, the
excitation and de-excitation will depend both on the acceleration and the distance to
the mirror. Although in this paper we have not presented an analysis of the transition amplitudes for the parallel motion, the presence of dissipative effects is clear from the computation of the imaginary part of the effective action. These effects have no analogs for a static atom in front of a
moving (perfect) mirror. 

\section*{Acknowledgments}
This research was supported by Agencia Nacional de Promoci\'on Cient\'ifica
y Tecnol\'ogica (ANPCyT), Consejo Nacional de Investigaciones Cient\'ificas
y T\'ecnicas (CONICET), Universidad de Buenos Aires (UBA) and Universidad
Nacional de Cuyo (UNCuyo). The research of F.C.L. was supported in part by 
the National Science Foundation under Grant No. PHY-1748958.


\begin{thebibliography}{99}
\bibitem{footnote} In \cite{PaperI}, there is a mistake in the captions of Figs. 1 and 2, where the labels for parallel and normal motions were unintentionally interchanged.
\end{thebibliography}
\end{document}